\documentclass[twoside]{dis09}
\usepackage[latin1]{inputenc}
\usepackage[dvips]{graphicx,epsfig,color}
\usepackage{wrapfig,rotating}
\usepackage{amssymb,amsmath,array}

\newcommand{\ifb}{\ensuremath{\mathrm{fb^{-1}}}}
\newcommand{\TeV}{\ensuremath{\mathrm{Te\kern -0.1em V}}}
\newcommand{\TeVc}{\ensuremath{\mathrm{Te\kern -0.1em V\!/}c}}
\newcommand{\TeVcc}{\ensuremath{\mathrm{Te\kern -0.1em V\!/}c^2}}
\newcommand{\GeV}{\ensuremath{\mathrm{Ge\kern -0.1em V}}}
\newcommand{\GeVc}{\ensuremath{\mathrm{Ge\kern -0.1em V\!/}c}}
\newcommand{\GeVcc}{\ensuremath{\mathrm{Ge\kern -0.1em V\!/}c^2}}
\newcommand{\MeV}{\ensuremath{\mathrm{Me\kern -0.1em V}}}
\newcommand{\MeVc}{\ensuremath{\mathrm{Me\kern -0.1em V\!/}c}}
\newcommand{\MeVcc}{\ensuremath{\mathrm{Me\kern -0.1em V\!/}c^2}}

\newcommand{\um}{\ensuremath{\mathrm{\mu m}}}

\newcommand{\babar}{\mbox{\slshape B\kern-0.1em{\small A}\kern-0.1em B\kern-0.1em{\small A\kern-0.2em R}}}

\def\mass{4143.0\pm2.9(\mathrm{stat})\pm1.2(\mathrm{syst})~\MeVcc }
\def\width{11.7^{+8.3}_{-5.0}(\mathrm{stat})\pm3.7(\mathrm{syst})~\MeVcc }
\def\yield{14\pm5 }
 
\def\massdifffit{1046.3\pm2.9~\MeVcc }
\def\widthfit{11.7^{+8.3}_{-5.0}~\MeVcc }

\begin{document}
\title{Properties of Exotic Charmonium-like States at CDF}

\author{Kai Yi$^1$ for the CDF Collaboration
%
%
\vspace{.3cm}\\
%
1- University of Iowa - Dept of Physics and Astronomy \\
Iowa City, IA 52242-1489
}

\maketitle

\begin{abstract}
We report the recent evidence for a new narrow structure, 
$Y(4140)$, decaying to the $J/\psi \phi$ final state, in exclusive 
$B^+\rightarrow J/\psi\phi K^+$ decays in a data sample corresponding to an 
integrated luminosity of 2.7 \ifb collected at the CDF II detector. This 
narrow structure with its mass well above open charm pairs is unlikely  to 
be a candidate for a  
conventional charmonium state. 
From a study of the $X(3872)$ mass and width based on the world's largest
sample of $X(3872)\rightarrow J/\psi\pi^+\pi^-$ decays, we find that our $X(3872)$ signal
is consistent with a single state, and leads to the most precise measurement 
of the $X(3872)$ mass.

\end{abstract}

\section{Introduction}

The existence of exotic mesons beyond $q\bar{q}$ has been discussed for 30 years~\cite{PDG},
but evidence for such mesons has not  been clearly established.
The recently discovered states   
that have charmonium-like decay modes~\cite{x3872discovery} 
but are difficult to place in the overall charmonium system
have introduced challenges to conventional $q\bar{q}$ meson model.  
The possible interpretations beyond $q\bar{q}$   
such as hybrid ($q\bar{q}g$) and four-quark states ($q\bar{q}q\bar{q}$)
have revitalized  interest in exotic mesons in the charm 
sector~\cite{conventional}.   
So far the observed states  involve 
only $c$ quark  and light quark ($u$, $d$) decay products,   
however,  the $J/\psi\phi$ final state  
is a good channel for an exotic meson search which extends to $c$ quark 
and heavy $s$ quark decay products. 
An investigation of the $J/\psi\phi$ system produced in exclusive 
$B^+\rightarrow J/\psi\phi K^+$ decays with $J/\psi \rightarrow \mu^+ \mu^-$ 
and $\phi\rightarrow K^+K^-$ is reported in this note. 
This analysis is based on a data sample  of $\bar{p} p $
collisions at $\sqrt{s}=1.96~\TeV$ with an integrated luminosity of 2.7 $\ifb$
collected by the CDF II detector at the Tevatron. 
Charge conjugate modes are included implicitly in this note.

\begin{wrapfigure}{r}{0.5\columnwidth}
\centerline{
\includegraphics[width=0.45\columnwidth]{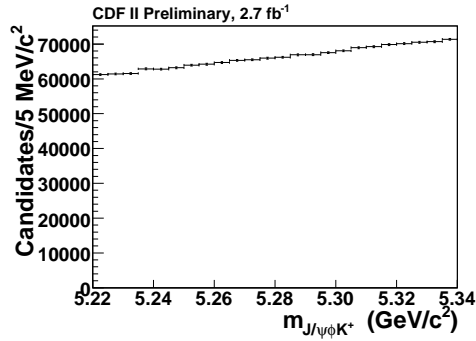}
}
\caption{
The  $J/\psi\phi K^+$ mass before   
minimum ${L_{xy}(B^+)}$ and  kaon $LLR$ requirements. 
}\label{Fig1}
\end{wrapfigure}

It has been six years since the discovery of $X(3872)$, however, the nature of this state 
has not been clearly understood yet. Due to the closeness of the $X(3872)$ mass with 
the threshold of $D^0$ and $D^{*0}$, the $X(3872)$ has been proposed as a 
molecule composed of $D^0$ and $D^{*0}$. The $X(3872)$ has also been speculated 
to be two states nearby, as in some models like the $diquark-antidiqurk$ model. 
It is critical to make precise measurements of the mass and width of 
$X(3872)$ to understand its nature.
The  large  $X(3872)\rightarrow J/\psi\pi^+\pi^-$ sample accumulated at CDF enables us 
to  test the hypothesis of that the $X(3872)$ is composed of two states and to make a 
 precise mass measurement of $X(3872)$ if it is consistent with one state hypothesis.

\section{Evidence of Y(4140) }

\begin{wrapfigure}{r}{0.5\columnwidth}
\centerline{
\includegraphics[width=0.45\columnwidth]{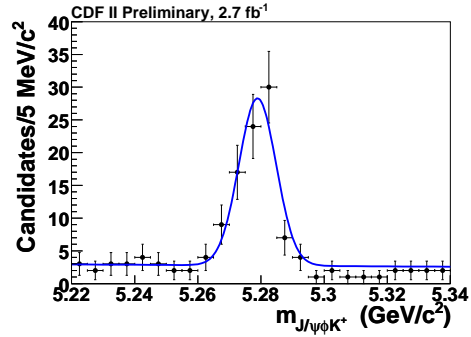}
}
\caption{
The $J/\psi\phi K^+$ mass after  
minimum ${L_{xy}(B^+)}$ and $LLR$ requirements; the solid line is a fit to 
the data with a Gaussian signal 
function and flat background function.
}\label{Fig2}
\end{wrapfigure}

We first reconstruct $B^+\rightarrow J/\psi\phi K^+$ signal and 
then search for structures in the $J/\psi\phi$ mass spectrum~\cite{y4140}.
The $J/\psi\rightarrow \mu^+\mu^-$ events are recorded using a dedicated dimuon trigger. 
The  $B^+\rightarrow J/\psi\phi K^+$ candidates are reconstructed 
by combining a $J/\psi\to\mu^+\mu^-$ candidate, a $\phi\rightarrow K^+K^-$ candidate,  
and an additional charged track. 
Each track  is required to have at least 4  axial silicon  hits and have a transverse momentum 
greater than 400 \MeVc. 
The reconstructed mass of each vector meson candidate must lie within 
a suitable range from the nominal 
values ($\pm$50 \MeVcc~ for the $J/\psi$ and $\pm$7 \MeVcc~ for the $\phi$).  In the final $B^+$ 
reconstruction the $J/\psi$ is mass constrained, and the $B^+$ candidates must have
$p_T > 4$ \GeVc.
The $P(\chi^2)$ of the mass- and vertex-constrained fit to the 
$B^+\rightarrow J/\psi\phi K^+$ candidate is required to be greater than 1\%.

\begin{wrapfigure}{r}{0.5\columnwidth}
\centerline{\includegraphics[width=0.45\columnwidth]{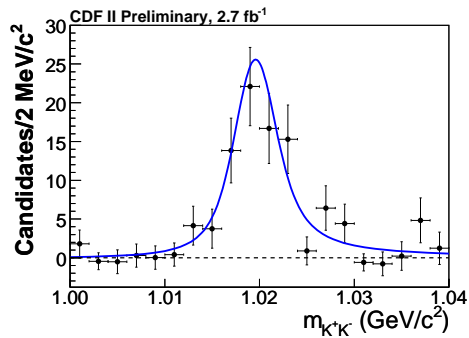}
}
\caption{
The $B^+$ sideband-subtracted   $K^+K^-$ mass without the $\phi$ mass window requirement.
The solid curve is a $P$-wave relativistic Breit-Wigner fit to the data. 
}\label{Fig3}
\end{wrapfigure}

To suppress combinatorial background, we use ${dE/dx}$ and 
$\mathrm{TOF}$ information to identify all three kaons in the final state.
The information is summarized in a log-likelihood ratio ($LLR$), which 
reflects how well a candidate track can be positively identified 
as a kaon relative to other hadrons.
In addition, we require a minimum ${L_{xy}(B^+)}$ for the 
$B^+\rightarrow J/\psi\phi K^+$ candidate,
where ${L_{xy}(B^+)}$ is the projection onto ${\vec{p}_T(B^+)}$ 
of the vector connecting the primary vertex to the ${B^+}$ decay vertex.   
The  ${L_{xy}(B^+)}$ and   ${LLR}$ requirements for $B^+\rightarrow J/\psi\phi K^+$ 
are then chosen to 
maximize $\mathit{S/\sqrt{S+B}}$,  
where $\mathit{S}$ is the number of $B^+\rightarrow J/\psi\phi K^+$ signal events 
and $\mathit{B}$ is the number of background events  implied from the 
$B^+$ sideband.  
The requirements  obtained by  maximizing $\mathit{S/\sqrt{S+B}}$ are ${L_{xy}(B^+)}>500 ~\um$ 
and  ${LLR}>0.2$.

The invariant mass of $J/\psi\phi K^+$, after $J/\psi$ and $\phi$ mass window  
requirements, and before and after the minimum ${L_{xy}(B^+)}$ and kaon ${LLR}$ requirements, 
are shown in Fig.~\ref{Fig1} and  Fig.~\ref{Fig2}, respectively. 
We do not see $B^+$ signal at 
all before the ${L_{xy}(B^+)}$ and kaon ${LLR}$ requirements, but we see clear 
$B^+$ signal after the  requirements. 
A fit with a Gaussian signal function and a 
flat background function  to the mass spectrum of $J/\psi\phi K^+$ (Fig.~\ref{Fig2}) 
returns a $B^+$ signal of $75\pm10(\mathrm{stat})$ events.
The ${L_{xy}(B^+)}$ and ${LLR}$  requirements reduce the background by a factor of 
approximately 20 000 
while keeping a signal efficiency of approximately 20\%.
We select  $B^+$ signal candidates with a mass 
within  3$\sigma$  of the nominal $B^+$ mass; 
the purity of the $B^+$ signal in that mass window is about 80\%.

\begin{wrapfigure}{r}{0.5\columnwidth}
\centerline{\includegraphics[width=0.45\columnwidth]{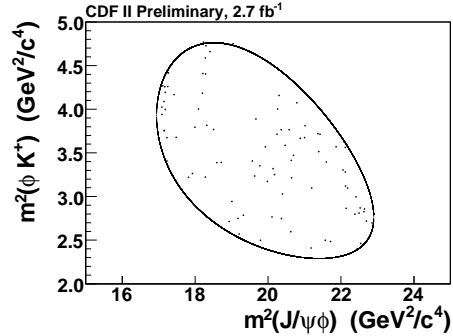}}
\caption{ The Dalitz plot of ${m^2(\phi K^+)}$ 
versus ${m^2(J/\psi \phi)}$  in the  $B^+$ 
mass window. The boundary shows the kinematic allowed region.}\label{Fig4}
\end{wrapfigure}

The combinatorial background under the $B^+$ peak includes
$B$ hadron decays such as $B^0_s \rightarrow \psi(2S)\phi \rightarrow J/\psi \pi^+\pi^-\phi$,
in which the pions are misidentified as kaons.  However,
background events with misidentified kaons cannot
yield a Gaussian peak at the $B^+$ mass consistent with the
5.9 \MeVcc mass resolution.  
Figure~\ref{Fig3} shows the  $K^+ K^-$ mass  
from $\mu^+\mu^-K^+K^- K^+$ candidates within $\pm 3 \sigma$ of the nominal $B^+$ mass
with $B$ sidebands subtracted and before applying $\phi$ mass window requirement.
Using a smeared $P$-wave relativistic Breit-Wigner (BW)~\cite{pbw} line-shape fit to the spectrum 
returns a $\chi^2$ probability of 28\%.  
This shows that 
the $B^+ \rightarrow J/\psi K^+K^-K^+$ final state is well described 
by $J/\psi \phi K^+$.

\begin{wrapfigure}{r}{0.5\columnwidth}
\centerline{\includegraphics[width=0.45\columnwidth]{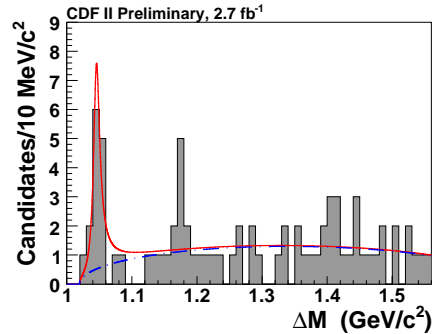}}
\caption{
 The mass difference, $\Delta M$, between $\mu^+\mu^-K^+K^-$ and $\mu^+\mu^-$,  
 in the $B^+$ mass window. 
The dash-dotted curve is 
the background contribution and the red solid curve is the total unbinned fit.}\label{Fig5}
\end{wrapfigure}

We then examine the effects of detector acceptance and selection
  requirements using 
$B^+ \rightarrow J/\psi \phi K^+$ MC events simulated  by a phase space distribution. 
The MC events are smoothly distributed in the Dalitz plot and 
in the $J/\psi \phi$ mass spectrum. No artifacts were observed from MC events. 
Figure~\ref{Fig4} shows the Dalitz plot of ${m^2(\phi K^+)}$ 
versus ${m^2(J/\psi \phi)}$,   
and  Fig.~\ref{Fig5} shows 
the mass difference, $\Delta M= m(\mu^+\mu^-K^+K^-)- m(\mu^+\mu^-)$, for events in 
the $B^+$ mass window in our data sample. 
We examine the enhancement in the $\Delta M$ spectrum just above $J/\psi\phi$ threshold. 
We exclude the high--mass part of the spectrum beyond $1.56$ \GeVcc~ to avoid 
combinatorial backgrounds that would be expected from misidentified 
$B^0_s\rightarrow \psi(2S) \phi\rightarrow (J/\psi\pi^+\pi^-)\phi$ decays.
We model the enhancement by an $S$-wave relativistic BW 
function~\cite{sbw} convoluted with 
a Gaussian resolution function with the RMS fixed to 1.7 \MeVcc~obtained from MC,  
and use three--body phase space~\cite{PDG} to describe the background shape.
An unbinned likelihood fit to the $\Delta M$ distribution,   
as shown in Fig.~\ref{Fig5}, returns a yield of $\yield$ events,   
a $\Delta M$ of $\massdifffit$, and a width of $\widthfit$.

We use the log-likelihood ratio of $-2{\mathrm{ln}}(\mathcal{L}_0/\mathcal{L}_{{max}})$
to determine the significance of the enhancement, 
where $\mathcal{L}_0$ and $\mathcal{L}_{{max}}$ are the likelihood 
values for the null  hypothesis fit and signal hypothesis fit.
The $\sqrt{-2{\mathrm{ln}}(\mathcal{L}_0/\mathcal{L}_{{max}})}$ value 
is 5.3 for a pure three--body phase space 
background shape assumption. 
  We generate $\Delta M$ spectra using the background distribution
  alone, and search for the most significant fluctuation with 
$\sqrt{-2{\mathrm{ln}}(\mathcal{L}_0/\mathcal{L}_{{max}})}\ge 5.3$  in
  each spectrum in the mass range of 1.02 to 1.56 \GeVcc, with widths
  in the range of 1.7 (detector resolution) to 120 \MeVcc~(ten times the observed width).

\begin{wrapfigure}{r}{0.5\columnwidth}
\centerline{\includegraphics[width=0.45\columnwidth]{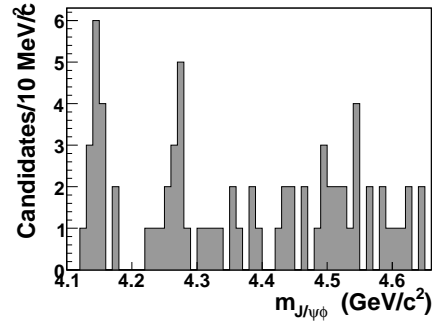}}
\caption{The $J/\psi\phi$ mass distribution  
 in the $B^+$ mass window.}\label{Fig6}
\end{wrapfigure}

The resulting $p$-value  from 3.1 million simulations 
is $9.3\times 10^{-6}$, corresponding to a significance of 4.3$\sigma$.
We repeat this process with a flat combinatorial non-B background and three--body PS 
for non-resonance $B$ background and we still get a significance of 3.8$\sigma$.

One's eye tends to draw to a second cluster of events
around 1.18 \GeVcc~in Fig.~\ref{Fig5}, or 
around 4.28 \GeVcc~in $J/\psi\phi$ mass as shown in Fig.~\ref{Fig6}.
It is  close to one pion mass
above the peak at the $J/\psi\phi$ threshold. However, this cluster 
is statistically insufficient to infer the presence of a second structure. 
To investigate possible reflections, we examine the Dalitz plot and 
projections into  $\phi K^+$ and $J/\psi K^+$ spectrum.  
We find no evidence for any other structure in the $\phi K^+$ and $J/\psi K^+$ 
spectrum.

\section{Measurement of the mass of $X(3872)$}

\begin{wrapfigure}{r}{0.5\columnwidth}
\centerline{\includegraphics[width=0.45\columnwidth]{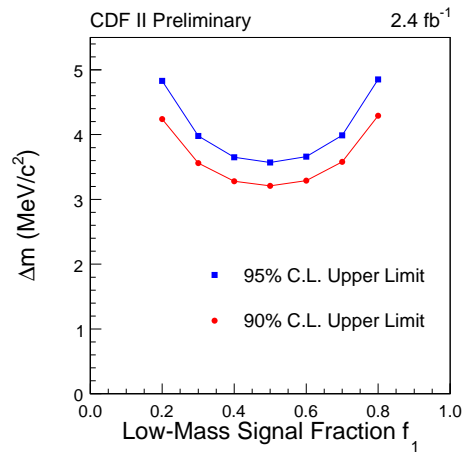}}
\caption{
Dependence of the upper limit on the mass difference $\Delta m$ between two states on the 
low-mass state signal fraction $f_1$.
}\label{Fig7}
\end{wrapfigure}

In studying our $X(3872)$ sample we tested the hypothesis
of whether the observed signal stems from two different states,
as predicted in four-quark models. We fit the 
$X(3872)$ mass signal with a BW function convoluted with a resolution function~\cite{x3872mass}. 
Both functions contain a width scale factor that is a free parameter in the fit 
and therefore sensitive to the shape of the mass signal. The measured width scale factor is 
compared to the values seen in pseudo experiments which assume two states with given mass 
difference and ratio of events. The resolution in the simulated events is corrected for the 
difference between data and simulation measured for the $\psi(2S)$.
The result of this hypotheses test shows that the data is completely consistent with  a single 
state. Under the assumption of two states with equal amount of observed events, we set a limit of
$\Delta m < 3.2 (3.6)$ \MeVcc at 90\% (95\%) C.L.
The limit for other ratios of events in the two peaks is shown in Fig.~\ref{Fig7}.

Since our signal is consistent with one peak, we proceed and measure the X(3872) mass in an 
unbinned maximum likelihood fit. The systematic uncertainties are determined from the difference 
between the measured $\psi(2S)$ mass and its world average value, the potential variation of the 
$\psi(2S)$ mass as a function of kinematic variables, and the difference in Q value between $X(3872)$  
and $\psi(2S)$. Systematic effects due to the fit model are negligible. The measured $X(3872)$ mass is:
$m(X(3872)) = 3871.61 \pm 0.16 (stat) \pm 0.19 (syst)$ \MeVcc, which  
is the most precise measurement to date as shown in Fig.~\ref{Fig8}.

\section{Summary}

\begin{wrapfigure}{r}{0.5\columnwidth}
\centerline{\includegraphics[width=0.45\columnwidth]{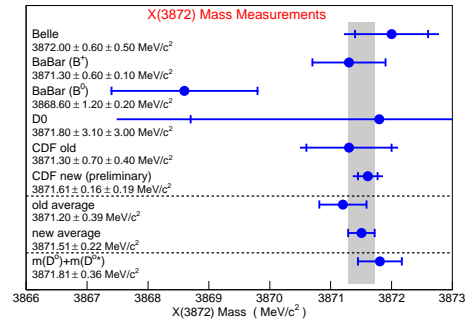}}
\caption{
An overview of the measured $X(3872)$ masses.
}\label{Fig8}
\end{wrapfigure}

The large  sample of
$B^+ \rightarrow J/\psi \phi K^+$ decays  enables us 
to search for structure in the $J/\psi\phi$ mass spectrum,  
and we find  evidence for a narrow structure near the $J/\psi\phi$ 
threshold with a significance estimated to be at least  3.8$\sigma$.
Assuming an $S$-wave relativistic BW, the mass (adding $J/\psi$ mass) and width of this structure, 
including systematic  uncertainties,  are measured to be $\mass$ 
and $\width$, respectively.   This structure does not 
fit conventional expectations for a charmonium state 
because it is expected to have a tiny BR to $J/\psi\phi$ with its  mass well beyond open charm pairs.
We term it the $Y(4140)$. The BR of $B^+\rightarrow Y(4140) K^+, Y(4140)\rightarrow 
J/\psi\phi $ is estimated to be $9.0\pm3.4(stat)\pm2.9(B_{BF}))\times10^{-6}$.

Studies using our $X(3872)$ sample, the largest in the world,
indicate that the $X(3872)$ is consistent with 
one state hypothesis and this leads to the most precise mass measurement of $(X3872)$ 
based on the one state hypothesis. 
The value is below, but within uncertainties of the $D^{*0}D^0$ threshold. The explanation 
of the X(3872) as a bound D*D system is therefore still an option.

\begin{footnotesize}


\end{footnotesize}


\begin{thebibliography}{99}




\bibitem{PDG}
C.~Amsler {\it et al.}  ~(Particle Data Group), Phys.\ Lett.\ B {\bf 667}, 1 (2008).



\bibitem{x3872discovery}
S.-K.~Choi {\em et al.}~(Belle Collaboration),  Phys.\ Rev.\ Lett. {\bf 91}, 262001 (2003);
D.~Acosta {\em et al.}~(CDF Collaboration), Phys.\ Rev.\ Lett. {\bf 93}, 072001 (2004);
S.-K.~Choi {\em et al.}~(Belle Collaboration), Phys.\ Rev.\ Lett. {\bf 94}, 182002 (2005);
B.~Aubert  {\em et al.}~(\babar~ Collaboration), Phys.\ Rev.\ Lett. {\bf 95}, 142001 (2005);
S.~Godfrey and S.~ L. ~Olsen,  Ann.\ Rev.\ Nucl.\ Part. \ Sci {\bf 58} (2008) 51.


\bibitem{conventional}
E.~Eichten, S.~Godfrey, H.~Mahlke, and J.~Rosner, Rev.\ Mod.\ Phys. {\bf 80}, 1161 (2008);
S.~L.~Zhu, Phys.\ Lett.\ B {\bf 625}, 212 (2005);
F.~Close  and P.~Page, Phys.\ Lett.\ B {\bf 628}, 215 (2005);
E.~S.~Swanson,  Phys.\ Lett.\ B {\bf 588},  189 (2004);
L.~Maiani, F.~Piccinini, A.~D.~Polosa, and V.~Riquer, Phys.\ Rev.\ D {\bf 72}, 031502(R) (2005); 
N.~V.~Drenska, R.~Faccini, and A.~D.~Polosa, 	arXiv:hep-ph/0902.2803v1.

\bibitem{y4140}
T.~Aaltonen {\em et al.} CDF Collaboration,  Phys.\ Rev.\ Lett. {\bf 102}, 242002 (2009).



\bibitem{pbw}
B.~Aubert  {\em et al.}~(\babar~ Collaboration), Phys.\ Rev.\ D {\bf 78},  071103 (2008).


\bibitem{sbw}
$\frac{dN}{dm} \propto \frac{m \Gamma(m)}{(m^2-m_0^2)^2+m_0^2\Gamma^2(m) }$, where
$\Gamma(m)=\Gamma_0\frac{q}{q_0}\frac{m_0}{m}$, and the 0 subscript indicates the 
value at the peak mass.


\bibitem{x3872mass}
CDF Collaboration, CDF public note 9454;
http://www-cdf.fnal.gov/physics/new/bottom/bottom.html.




\end{thebibliography}
\end{document}